\documentclass{iau}

\usepackage{graphicx} 

\title[IAUS291.~~Magnetar QPOs: variability and emission] 
{Quasi-Periodic Oscillations in magnetars: linking variability and emission} 

\author[C. D'Angelo] 
{Caroline D'Angelo}

\affiliation{Instituut Anton Pannekoek, University
    of Amsterdam, Amsterdam, Netherlands 1098 XH \\
email: {\tt  c.r.dangelo@uva.nl}
}

\pubyear{2012}
\volume{291}  
\jname{\mbox{Neutron Stars and Pulsars: Challenges and Opportunities after 80 years}}
\editors{J. van Leeuwen, ed.} 
\begin{document}

\maketitle

\begin{abstract}
  I present recent results studying flare emission in
  magnetars. Strong quasi-periodic oscillations observed in the tail
  of giant magnetar flares are frequently interpreted as evidence for
  global seismic oscillations. I demonstrate that such a global
  oscillation is not directly observable in the lightcurve. New work
  suggests the amplitude for the strongest QPO stays nearly constant
  in the rotation phases where it is observed, which I argue suggests
  it is produced by an additional emission process from the star.

\end{abstract}

\firstsection 
\section{Introduction}
In the past few years, considerably theoretical interest has been
focused on the seismology of neutron stars (e.g. Levin 2006;
Glampedakis et al. 2006; Gabler et al. 2011), largely been motivated
by the observation of a series of quasi-periodic oscillations (QPOs)
in the decay tails of the giant flares in magnetars SGR 1900+14 and
SGR 1806-20 (Israel et al. 2005; Watts \& Strohmayer 2006; Strohmayer
\& Watts 2006).  The large amount of energy released during a flare
($\sim 10^{46}~$erg) is likely powered by a global reconfiguration of
the star's magnetic field (Thompson \& Duncan 1995) and some of that
energy will trigger large-scale quakes in the star (Duncan 1998).

Comparatively little work has been done to connect the putative
starquakes directly to the observed QPOs. Timokhin et al. (2008)
recently proposed that the oscillations could be attributed to a
variable current density in the stellar magnetosphere, created by
twisted magnetic field lines anchored in the vibrating star. These
electrons then Compton upscatter photons from the surface of the star,
modifying the observed spectrum.

In this article I revisit the properties of the QPOs observed in SGR
1806-20 and argue that correlations (or lack thereof) between the
variability on different timescales can be used to put strong
constraints not only on the QPO origins but also on the emission
mechanisms powering the flares themselves.

\section{Quasi-Periodic Oscillations in SGR 1806-20}

Six distinct QPOs were detected in the SGR 1806-20 giant flare using
X-ray data from both the $RHESSI$ and $RXTE$ satellites. The QPOs have
central frequencies between 17 and 1837 Hz, with fractional rms
amplitudes between 4 and 20\%. They are further characterized by a
high degree of coherence: of the six detected QPOs, only the one at
150 Hz has a width (full-width at half maximum) of 17 Hz; the others
have widths between 1 and 5 Hz (Strohmayer \& Watts 2006). There is
some evidence for energy dependence in the QPO at 625 Hz, which had an
rms amplitude of $\sim8\%$ below 100 keV, but of rms $\sim20\%$ between
100-200 keV (Watts \& Strohmayer 2006). The energy dependence of the
other QPOs is not clearly detected, but cannot be excluded due to
uncertainties in the measured photon energies.

Different QPOs were detected at different time intervals in the decay
tail of the flare and at different phases in the rotational pulse
profile. The majority of the QPOs were strongest beginning $\sim200$s
after the main flare, and in the `interpulse' region in phase, where
the flux in the lightcurve is at a minimum (see
fig. \ref{fig:QPO_intensity} for the the averaged pulse profile). The
strongest QPO in the RXTE data has a central frequency of 93 Hz and
fractional rms amplitude of $\sim 20\%$. It is also detectable over a
significant portion of the rotation phase and hence can be used to
study the relationship between the QPOs and the pulse profile, as well
as relationships between the QPO and broadband noise.

\section{Direct detection of a starquake}

D'Angelo \& Watts (2012) studied whether a starquake could have an
observable effect directly on the lightcurve itself by shaking the
emitting region. If the pulse profile is very steep (i.e. some
component of the pulse is beamed) then the sharp edge of the beam will
amplify the underlying motion of the surface, much like a flashlight
wiggling in and out of an observer's line of sight. The change to the
rotational phase from the crust motion is given by:
\begin{equation}
\label{eq:delPhi}
\Delta \Phi=\frac{\Delta x}{R_*\sin i\sin\alpha}
\sin(2\pi\nu_0t),
\end{equation}
where $\frac{\Delta x}{R_*}$ is the fractional amplitude of the
starquake, $\nu_0$ is its frequency, and $\sin i$ and $\sin \alpha$
are geometrical factors depending on the beam orientation.

D'Angelo \& Watts (2012) found that although significant amplification
of a star quake is possible, for the observed lightcurves and
realistic, the effect can be excluded. The fractional rms amplitude of
a QPO with phase change $\Delta \Phi$ is given by:
\begin{equation}
\label{eq:amplitude}
A\sim\frac{\frac{dP}{d\Phi}\Delta\Phi}{\langle P{\Phi}\rangle},
\end{equation}
where $P(\Phi)$ is the pulse profile as a function of rotation
phase. The amplification provided by a steep gradient is not enough to
make a starquake (with $\Delta x/R_* < 0.01$) directly detectable. This
result also excludes the possibility that weak, extremely steep
`pencil beams' (unresolved in the lightcurve) can provide the
amplification. The amplification factor in that case will be given by
eq. \ref{eq:amplitude} times an additional factor $P_{beam}/\langle
P\rangle$, the fractional amplitude of the steep beam. The beam
gradients required in this case are steep enough to be plausibly
excluded.

This result strongly suggests the QPOs are produced by variations in
the amplitude of the emission itself, rather than the starquake
directly.
\section{Modulation versus Emission}
The physical properties of the QPO can be somewhat constrained from
the observed variability of the lightcurve. This is most easily seen
from the power spectrum, the squared amplitude of the Fourier
transform of a segment of the lightcurve (e.g. van der Klis 1989). The
left and right panels of figure \ref{fig:QPO_intensity} show power
spectra from the pulsed tail of the giant flare, centered at two
different phases of the pulse profile (shown in the bottom panel). In
each power spectrum the QPO at 93 Hz is clearly visible, and a fit to
the QPO is overlaid. The QPO is significantly narrower in the
interpulse region, and the broadband noise (below $\sim100$ Hz) is
lower (a second QPO is seen at 30 Hz in the interpulse spectrum). At
the same time, the mean flux in the lightcurve is $\sim60\%$ lower
than in the secondary pulse.

There are two obvious ways that the observed intensity can vary at the
QPO frequencies. Either the surface flux can be modulated by a
quasi-periodically varying process (like changing optical depth to
electron scattering, cf. Timohkin et al. 2008), or the amount of flux
being emitted by the star can vary, either through variations in the
overall surface emission or via some MHD instability that produces
emission. At present the idea of variable emission in the magnetar
magnetosphere is purely speculative, but mechanisms for producing QPOs
in solar flares are an active research topic, and some of these could
potentially be relevant for magnetar flares as well (see e.g. the
review by Nakariakov \& Melnikov 2009).

\begin{figure}[t]
\begin{center}
 \includegraphics[width=3.8in]{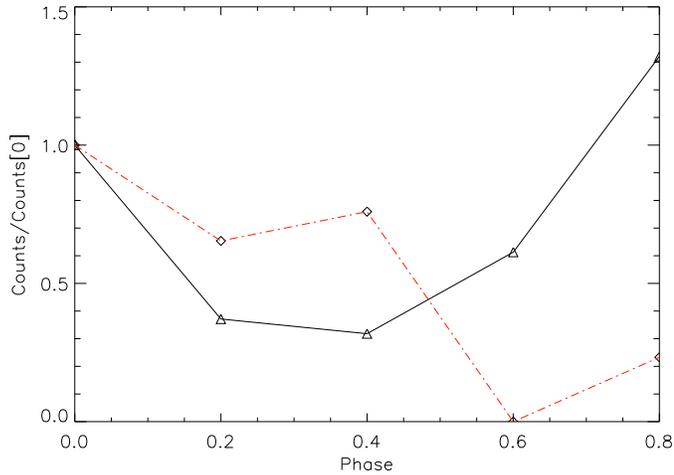} 
 \caption{Mean flux (black solid line) and 93Hz QPO amplitude (red
   dash-dotted line) as a function of phase, for the tail of the SGR
   1806-20 giant flare. For the phases where the QPO is detected, the
   amplitude is consistent with being constant while the mean flux
   varies substantially.}
   \label{fig:QPO_phase}
\end{center}
\end{figure}

The difference between an emitting process and a modulating one should
be observable in the phase-resolved QPOs. A modulating process should
produce a correlation between the absolute amplitude of the QPO and
the mean flux. In contrast, an emission process should stay constant
in phase, and be stronger at phases when the mean flux is
lower. Figure \ref{fig:QPO_phase} shows the fractional change in flux
as a function of phase (black line solid), overplotted with the
fractional change in QPO amplitude (integrated power over 20 Hz
centered at 93 Hz; red dot-dashed line). As is evident from the figure,
the QPO amplitude does not drop as much in the interpulse region, then
disappears altogether at the main peak of the burst. The lack of
correlation between mean flux and QPO amplitude would suggest that an
additional emission process is responsible for its production.

\section{Markov Chain Monte Carlo Simulations}

\begin{figure}[tb]
\begin{minipage}[b]{0.5\linewidth}
\centering
\includegraphics[width=\textwidth]{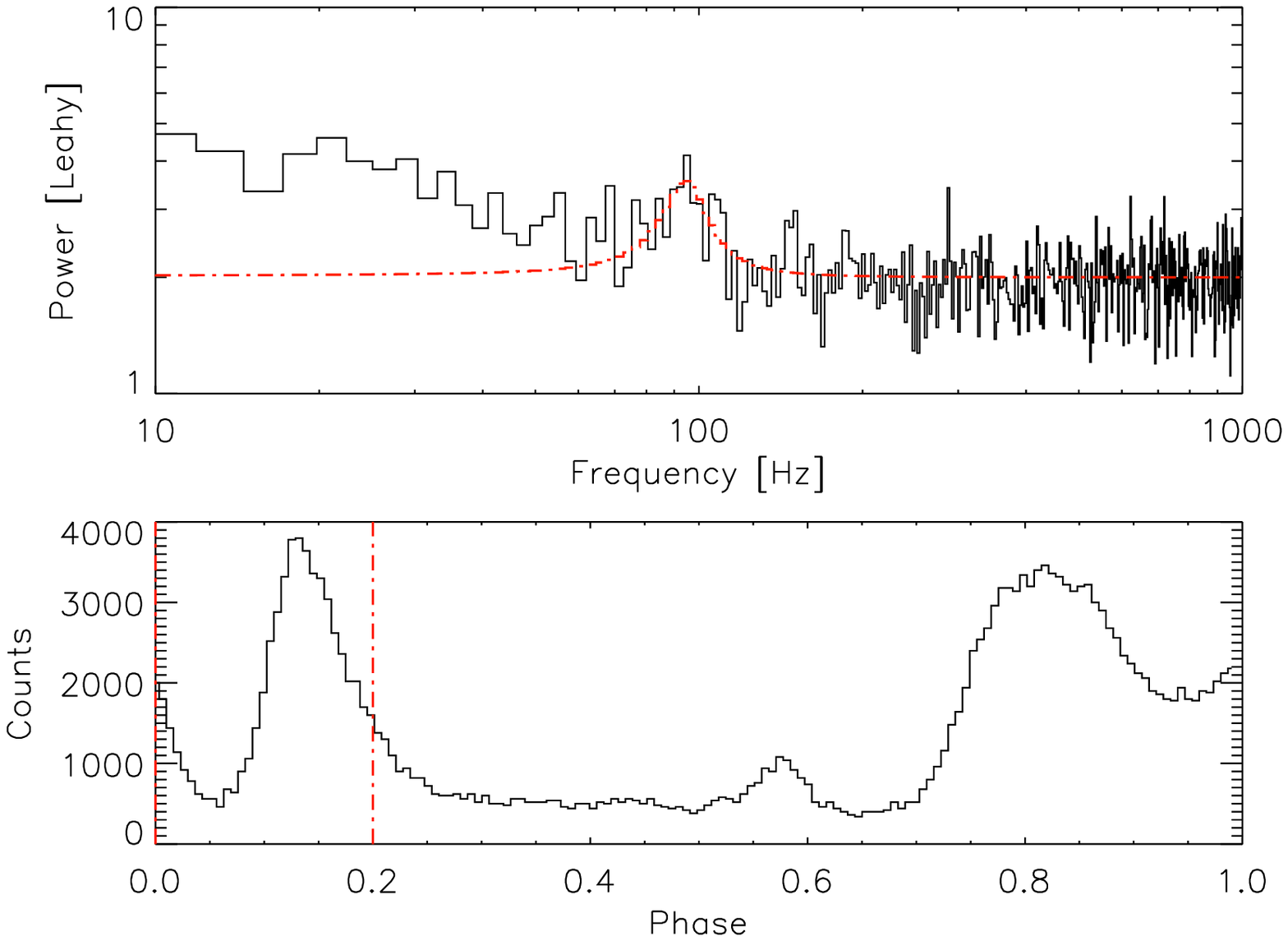}
\end{minipage}
\begin{minipage}[b]{0.5\linewidth}
\centering
\includegraphics[width=\textwidth]{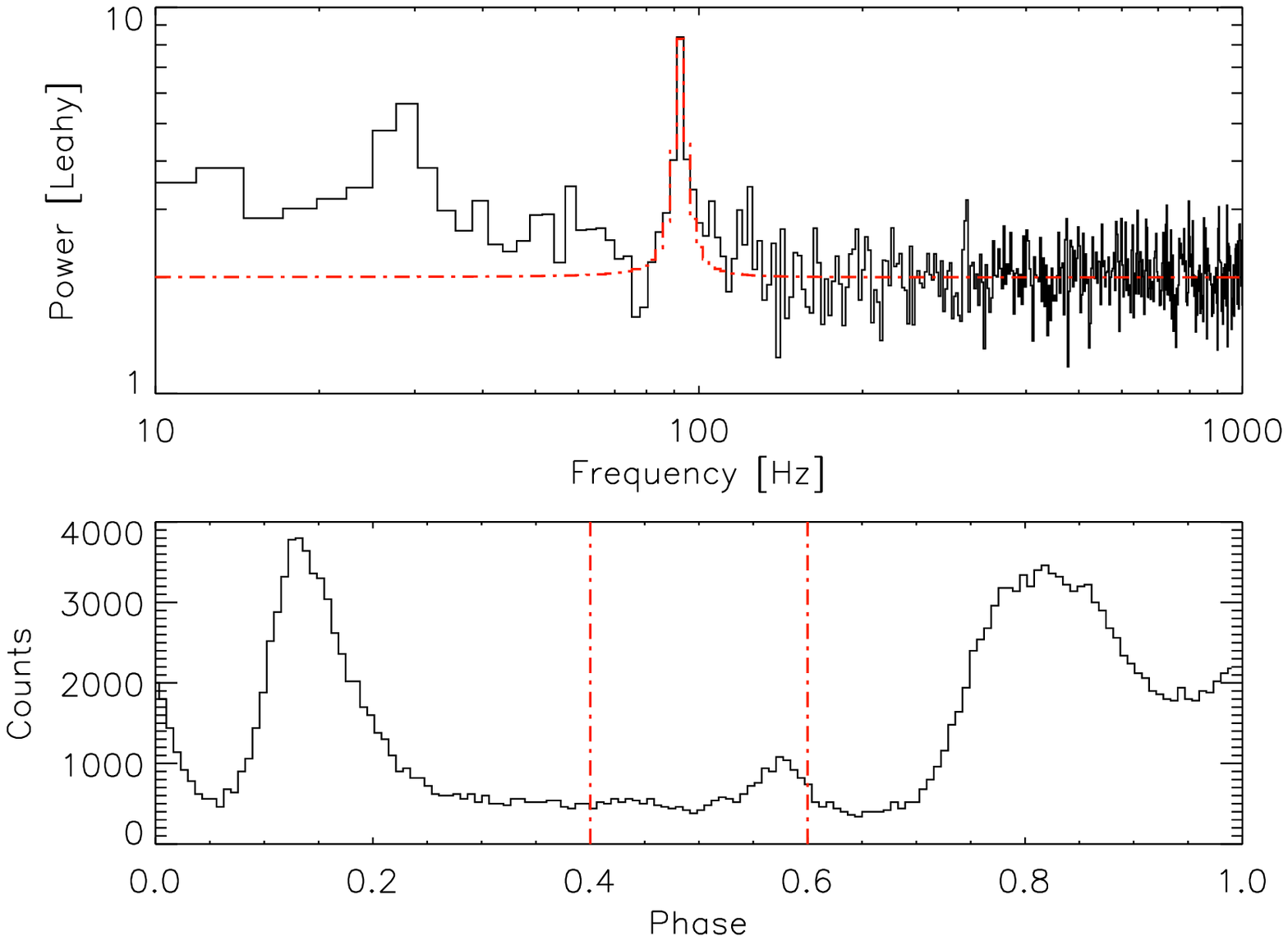}
\end{minipage}
\caption{Power spectrum and pulse profile of the tail of the SGR
  1806-20 giant flare for two different segments of the rotational
  phase (shown by the vertical lines). The QPO at 93 Hz is fit with a
  Lorentzian distribution (overlaid in red dash-dotted line).}
\label{fig:QPO_intensity}
\end{figure}

Determining the amplitude of the QPOs -- particularly those below
100Hz -- is complicated by the presence of low-frequency broadband
noise (visible in the left panel of
fig. \ref{fig:QPO_intensity}). Part of the signal at 93 Hz could
originate from red noise and not the QPO process, but disentangling
the two components is not straightforward (see e.g. Vaughan 2005).

To quantify the uncertainty in QPO amplitde, we use Markov Chain Monte
Carlo simulations to generate a series of realizations of a power
spectrum with a broadband component and a QPO given by the best fit to
the observed spectrum (Vaughan 2010). The variation in the resulting
measured parameters can be used to constrain the uncertainty on the
QPO fit, and more accurately determine the variation in QPO amplitude
as a function of phase. Preliminary results of this analysis suggest
that the amplitude observed QPO at 93 Hz is consistent with remaining
constant over the rotation phase where it is detectable. This would
seem to suggest it is independent of the secondary pulse peak, and
point to an underlying emission process. The definitive results will
be published in a forthcoming paper.

\section{Conclusions}
The variability of magnetar giant flares on different timescales shows
correlations that can constrain the underlying emission mechanism,
both for the QPOs and potentially the emission from the giant flares
themselves. We have excluded the possibility of directly detecting
surface oscillation of the magnetar crust, and have presented
preliminary evidence that suggests an additional emission mechanism
might be active to produce the quasi-periodic oscillations. 


\end{document}